\documentclass[aps,twocolumn,prb]{revtex4}

\usepackage{graphicx}% Include figure files
\usepackage{dcolumn}% Align table columns on decimal point
\usepackage{amsmath,amssymb,bbold,bm,epsfig}

\begin{document}

\newcommand{\cA}{{\cal A}}
\newcommand{\bj}{{\bf j}}
\newcommand{\bk}{{\bf k}}
\newcommand{\bp}{{\bf p}}
\newcommand{\bv}{{\bf v}}
\newcommand{\bq}{{\bf q}}
\newcommand{\tbq}{\tilde{\bf q}}
\newcommand{\tq}{\tilde{q}}
\newcommand{\bQ}{{\bf Q}}
\newcommand{\br}{{\bf r}}
\newcommand{\bR}{{\bf R}}
\newcommand{\bB}{{\bf B}}
\newcommand{\bE}{{\bf E}}
\newcommand{\bA}{{\bf A}}
\newcommand{\bK}{{\bf K}}
\newcommand{\vd}{{v_\Delta}}
\newcommand{\tr}{{\rm Tr}}
\newcommand{\bGa}{{\bm \Gamma}}
\def\id{\mathbb{1}}

\title{Witten effect in a
crystalline topological insulator}
\author{G. Rosenberg}
\affiliation{Department of Physics and Astronomy,
University of British Columbia, Vancouver, BC, Canada V6T 1Z1}
\author{M. Franz}
\affiliation{Department of Physics and Astronomy,
University of British Columbia, Vancouver, BC, Canada V6T 1Z1}
\date{\today}

\begin{abstract}
It has been noted a long time ago that a term of the form $\theta (e^2/2\pi h)\bB\cdot\bE$ may be added to the
standard Maxwell Lagrangian without modifying the familiar laws of
electricity and magnetism. $\theta$ is known to particle physicists as
the `axion' field and whether or not it has a nonzero expectation
value in vacuum remains a fundamental open question of the Standard
Model. A key manifestation of the axion term is the Witten effect: a
unit magnetic monopole placed inside a medium with $\theta\neq 0$ is
predicted to bind a (generally fractional) electric charge
$-e(\theta/2\pi+n)$ with $n$ integer. Here we conduct a first test of
the Witten effect based on the recently established
fact that the axion term  with
$\theta=\pi$ emerges naturally in the description of the
electromagnetic response of a new class of crystalline solids called
topological insulators -- materials distinguished by strong spin-orbit
coupling and non-trivial band structures. Using a simple physical model
for a topological insulator we demonstrate the existence of a
fractional charge bound to a monopole by an explicit numerical
calculation. We also propose a scheme for generating an `artificial'
magnetic monopole in a topological insulator film, that may be used to
facilitate the first experimental test of Witten's prediction.

\end{abstract}
\maketitle

\section{Axions}

The idea of the axion was introduced in 1977 by Peccei and
Quinn\cite{pq} as a means to solve what is known as the `strong CP
problem' in the physics of strong interactions. The strong CP problem, 
the details of which are quite subtle, 
has to do with the vacuum structure of Quantum Chromodynamics. 
In simple physical terms it can be
stated as a question: Why is the electric dipole moment of the neutron
(currently unobserved) so small? The Standard Model predicts a value for
the neutron dipole moment $|{\bf d}_n|\sim 10^{-16} \theta$ e cm, with
$\theta$ of order unity, that should be readily
measurable. Peccei-Quinn's solution promotes $\theta$ to a dynamical
field describing a new elementary particle, the axion, whose vacuum
expectation value has relaxed to a very small value, explaining the
smallness of $|{\bf d}_n|$. The actual value of $\theta$, and the
validity of the Peccei-Quinn solution and its
variants\cite{weinberg0,wilczek0} remain open questions of
considerable importance to fundamental physics. The axion is also believed to
be a viable candidate for the elusive dark matter that
comprises the majority of matter in our universe \cite{duffy1} and is
subject to active experimental searches.\cite{admx,axionbook}

In a remarkable development axion electrodynamics has recently emerged
as a key tool in the description of crystalline solids called strong
topological insulators (STIs). These three-dimensional time-reversal
invariant (TRI) materials possess anomalous band structures
characterized by a Z$_2$-valued topological
invariant.\cite{mele1,moore1} This invariant, called $\nu_0$, counts
the number of topologically protected gapless surface states (modulo 2). 
A non-zero invariant means that the surface of such an insulator will be
metallic. This behavior has been predicted to occur \cite{fu2,zhang1}
and subsequently experimentally discovered \cite{cava1,cava2,chen1} in
several 3-dimensional solids such as Bi$_{1-x}$Sb$_x$ alloys and
Bi$_2$Se$_3$, Bi$_2$Te$_3$ crystals. More recently it has been
realized,\cite{qi1,essin1} remarkably, that the electromagnetic
response of a STI is characterized by the axion term 
\begin{equation}\label{ax}
\Delta{\cal L}_{\rm axion}=\theta \left({e^2\over 2\pi h}\right)\bB\cdot\bE
\end{equation}
with $\theta=\pi$, the only non-zero value permitted
by the time-reversal symmetry. When the time-reversal symmetry is
broken, e.g.\ in a crystal showing weak magnetism, $\theta$ can
acquire an arbitrary value. Fluctuations in the magnetic order
parameter then act as a dynamical axion field and can be thought of as
emergent axion particles.\cite{li1} Thus, aside from possible
practical applications, crystalline solids with topologically
non-trivial band structures have the potential to provide tabletop
laboratories for the testing and exploration of fundamental physical
paradigms.

A fundamental property of the axion medium is the Witten
effect:\cite{witten1} in the quantum theory, a magnetic monopole of unit
strength (i.e.\ projecting magnetic flux $\Phi_0=hc/e$) immersed in
an axion medium must carry electric charge $-e(\theta/2\pi+n)$ with
$n$ integer. This effect, although theoretically well established, has
never been experimentally tested because until now both a suitable
axion medium and the means to produce a magnetic monopole have been
lacking. In this study we demonstrate how the connection between the
axion response\cite{qi1,essin1} and strong topological
insulators\cite{mele1,moore1,fu2,zhang1,cava1,cava2,chen1} may serve to
overcome both obstacles. We remark that a 1D realization of the Witten effect in antiferromagnetic spin chains was proposed a long time ago.\cite {affleck1} Here we furnish the first concrete physical example of the Witten effect in 3D by modeling a STI with a magnetic monopole
inserted in its bulk. We show that the monopole binds a fractional
charge $\pm e/2$ consistent with Witten's prediction.\cite{witten1} We
then discuss possible ways to overcome the second obstacle by creating
an {\em emergent} magnetic monopole in a topological insulator. This
can be achieved by exploiting the degrees of freedom associated with a
vortex in the exciton condensate that may emerge in a thin film
topological insulator under external bias.\cite{seradjeh1} We conclude
that the prospects for experimental verification of the Witten effect in a
tabletop experiment using a STI appear promising.

%%%%%%%%%%%%%%%%%%%%%%%%%%%%%%%%%%%%%%%%%%%%%%%
\section{Monopole and the Witten effect}

We start with a brief overview of the Witten effect. Although the effect
is quantum-mechanical in nature its essence can be understood by
studying the classical Maxwell's equations modified in the presence of
$\Delta{\cal L}_{\rm axion}$. The axion term revises both Gauss'
law and Amp{\`e}re's law by adding extra source
terms,\cite{wilczek2}
\begin{eqnarray} \nabla\cdot\bE &=& \rho -{\alpha\over 4\pi^2}
\nabla\theta\cdot\bB,
\label{m1} \label{GaussLaw} \\ 
\nabla\times\bB &=& {\partial\bE\over\partial t} +\bj +
{\alpha\over 4\pi^2} \left(\nabla\theta\times\bE +
{\partial\theta\over\partial t} \bB\right), \label{m2}
\end{eqnarray}
where $\alpha=e^2/\hbar c$ is the fine structure constant and $\rho$,
$\bj$ are the electric charge density and current, respectively. We
observe that for uniform, constant $\theta$, Eqs.\ (\ref{m1}) and
(\ref{m2}) revert to the familiar Maxwell's equations, consistent with
the notion that $\Delta{\cal L}_{\rm axion}$ can be written as a total
derivative in this case. An important related property\cite{wilczek2}
is the periodicity under $\theta\to\theta +2\pi n$ of the axion action
${\cal S}_{\rm axion}$, implying that $\theta$ can be chosen from
the interval $[0,2\pi)$.

Now consider a unit monopole, ${\nabla\cdot\bB=\Phi_0\delta(\br)}$,
placed at the origin, in a medium initially characterized by ${\theta=0}$. 
We wish to understand what happens when we turn on $\theta$ 
as a function of time (but keep it uniform in space). To this end we
set $\nabla\theta=0$ and $\bj=0$ (no currents in vacuum) and take
the divergence of Eq.\ (\ref{m2}) to obtain
\begin{equation}\label{m3} \nabla\cdot{\partial\bE\over\partial
t}+{\alpha\over 4\pi^2}{\partial\theta\over\partial t}\nabla\cdot\bB =
0.
\end{equation}
We see that an electric field is generated in this process.  Integrating
Eq.\ (\ref{m3}) over space and time, we find that this field can be
thought of as originating from a point electric charge $Q$ located at
the origin with magnitude
\begin{equation}\label{m4} Q=-{\alpha\over
4\pi^2}\Phi_0\Delta\theta=-{\Delta\theta\over 2\pi}e,
\end{equation}
where $\Delta\theta$ is the net change in $\theta$ and we assumed that 
there was no initial electrical charge bound to the monopole, as
should be the case for a charge-conjugation and parity (CP) invariant
theory\cite{witten1} with $\theta=0$.  In a topological insulator
$\Delta\theta=\pi$, thus one expects a magnetic monopole to bind
fractional charge
\begin{equation}\label{q}
Q=-e \left( {1\over 2}+n \right).
\end{equation}
The integer $n$ accounts for the possibility of binding extra electrons,
which can always occur -- only the fractional part of $Q$ is
non-trivial.

%%%%%%%%%%%%%%%%%%%%%%%%%%%%%%%%%%%%%%%%%%%%%%%
\section{Topological insulator as an axion medium}

We now specify our model for a topological insulator and show that it
indeed possesses the axion term. In order to minimize computational
difficulties we consider a very simple model, inspired by Ref.\
\onlinecite{fradkin1}, with electrons hopping on the cubic lattice
with two orbitals per site, denoted as $c$ and $d$. The Hamiltonian 
$H=H_{\rm SO}+H_{cd}$, consists of a spin dependent part, with hopping 
between neighboring sites of the lattice,
\begin{equation}\label{hso} H_{\rm
SO}=i\lambda\sum_{j,\mu}\Psi_{j}^\dagger\tau_z\sigma_\mu\Psi_{j+\mu} +
{\rm h.c.},
\end{equation}
where
$\Psi_j=(c_{j\uparrow},c_{j\downarrow},d_{j\uparrow},d_{j\downarrow})^T$,
$j$ labels sites of the cubic lattice, $\tau_\mu$ and $\sigma_\mu$ are Pauli
matrices in orbital and spin space, respectively, $\mu=x,y,z$, and the
spin-independent terms that connect the two orbitals,
\begin{equation}\label{hcd}
H_{cd}=\epsilon\sum_{j}\Psi_{j}^\dagger\tau_x\Psi_{j} -t\sum_{\langle
ij\rangle}\Psi_{i}^\dagger\tau_x\Psi_{j}+ {\rm h.c.}
\end{equation}
Although the model specified by Eqs.\ (\ref{hso}) and (\ref{hcd})
probably does not describe any real solid, it is physical in that it is
local in space and preserves time reversal and inversion
symmetries. We show below that for a range of parameters
$(\lambda,\epsilon,t)$ it represents a topological insulator, 
and therefore can be adiabatically deformed into any of the more
realistic models\cite{fu2,zhang1,guo1,pesin1} characterized by the same
topological invariants. Any physical property that depends only on the
topological invariants, such as the electrical charge bound to a
monopole, can thus be calculated in the present model and the result
will remain applicable to any topological insulator in the same
topological class. We note that a similar model was used in
Refs.\ \onlinecite{zhang1,li1,hosur1}.

Our Hamiltonian has a simple representation in momentum space,
$H=\sum_\bk\Psi_\bk^\dagger{\cal H}_\bk\Psi_\bk$ with
\begin{equation}\label{hk} {\cal
H}_\bk=-2\lambda\sum_\mu\tau_z\sigma_\mu\sin{k_\mu}+\tau_xm_\bk,
\end{equation}
and $m_\bk=\epsilon-2t\sum_\mu\cos{k_\mu}$. The spectrum of
excitations has two doubly degenerate bands,
\begin{equation}\label{ek}
E_\bk=\pm\sqrt{4\lambda^2(\sin^2{k_x}+\sin^2{k_y}+\sin^2{k_z})+m_\bk^2}.
\end{equation}
In the limit $\epsilon, t\to 0$ the bands touch at 8 non-equivalent
Dirac points located at $\bGa_{\ell=(n_xn_yn_z)}=\pi(n_x,n_y,n_z)$
with $n_\mu=0,1$. These $\bGa_{\ell}$ also coincide with the 8
time-reversal invariant momenta\cite{fu2} (TRIM). When $\epsilon, t$ are small
but non-zero, the low-energy excitations of the system can be
described in terms of 8 massive Dirac Hamiltonians
\begin{equation}\label{hdir} {\cal H}_\bk^\ell
=\sum_\mu\tau_z\sigma_\mu v_\mu^\ell k_\mu+\tau_x m_\ell
\end{equation}
obtained by a straightforward expansion of ${\cal H}_\bk$ to linear
order in momentum in the vicinity of $\bGa_{\ell}$.  Here
${v_\mu^\ell\equiv-2\lambda (-1)^{n_\mu}}$ are the Cartesian components of
the Dirac velocities at $\bGa_{\ell}$ and $m_\ell\equiv
m_{\bGa_{\ell}}$ are the corresponding Dirac masses. The Hamiltonians
${\cal H}_\bk^\ell$ show spectra
\begin{equation}\label{edir}
{E}_\bk^\ell=\pm\sqrt{4\lambda^2k^2+m_{\ell}^2}.
\end{equation}

The system described by Hamiltonian (\ref{hk}) is inversion symmetric
and we can thus employ the method devised in Ref.\ \onlinecite{fu2} to
determine the topological class of its insulating phases when the
negative-energy bands are filled with electrons. This straightforward
method requires computing the eigenvalues of the parity operator at
the 8 TRIM for the occupied bands. We find four distinct phases,
depending on parameters $\epsilon$ and $t$, two of which are STI. The
complete results are listed in Table I.
\begin{table}
\caption{$Z_2$ indices $(\nu_0;\nu_1\nu_2\nu_3)$ calculated according
to Ref.\ \onlinecite{fu2} for our model insulator at half filling and the
corresponding values of the axion parameter $\theta$. $\nu_0$ is the
important `strong' invariant while $\nu_{i=1,2,3}$ are the so called
`weak' invariants\cite{mele1,moore1} which do not play a role in the
present study but we list them here for completeness. It is assumed
that $\lambda,t>0$. WTI denotes a `weak' topological
insulator.} \label{z2table}

\begin{tabular}{ c c c c c }
  \hline

\ \ \  Parameters \ \ \ \ & \ \ \  $Z_2$ class \ \  & \ \ Insulator type \ \  & \ \ Axion  $\theta$ \ \  \\
  \hline
  $|\epsilon|> 6t$       & (0;000) & {\rm trivial} & 0 \\
  $-6t < \epsilon <-2t$  & (1;111) & {\rm STI}     & $\pi$ \\
  $-2t < \epsilon <2t$   & (0;111) & {\rm WTI}     & 0 \\
  $2t < \epsilon < 6t$   & (1;000) & {\rm STI}     & $\pi$ \\
  \hline
 \end{tabular}

\end{table}

According to general considerations\cite{qi1,essin1} the STI
phases should exhibit the axion term with $\theta=\pi$. We evaluate $\theta$
for our model using the non-abelian Berry
connection $\cA^{\alpha\beta}_i=-i\langle \alpha \bk|\partial_i
|\beta\bk\rangle$ and the formula\cite{qi1,essin1}
\begin{equation}\label{theta} \theta={1\over 4\pi}\int_{\rm
BZ}d^3k\epsilon^{ijk}\tr\left[\cA_i\partial_j \cA_k+ {2i\over
3}\cA_i\cA_j\cA_k\right],
\end{equation}
where $|\beta\bk\rangle$ is an eigenstate of ${\cal H}_\bk$, the trace
extends over occupied states and $\partial_i\equiv\partial/\partial
k_i$. The integral indicated in Eq.\ (\ref{theta}) is generally
difficult to evaluate and numerical methods must be used to obtain
$\theta$ for an arbitrary band structure. For a model in which the band
touching is described by Dirac Hamiltonians, however, a simple
analytical evaluation of Eq.\ (\ref{theta}) is possible by noticing
that in the limit of a small Dirac mass the entire contribution to the
integral comes from the Dirac points. We show in the Appendix that each
Dirac point contributes
\begin{equation}\label{thetai} \theta^\ell =-{\pi\over 2}{\rm
sgn}\left(v_x^\ell v_y^\ell v_z^\ell m_\ell\right)
\end{equation}
to the total $\theta=\sum_\ell\theta^\ell\mod 2\pi$. Although Eq.\
(\ref{thetai}) has been derived for the case $|m_\ell| \ll
|\lambda|$ we expect it to be more generally valid for all TRI
Hamiltonians that can be deformed into the form of Eq.\ (\ref{hk}). This
is because the value of $\theta$ in a TRI insulator is quantized and can
only change when a band crossing closes the gap. As long as band
crossings occur only at $\bGa_{\ell}$ points and are described by
Dirac Hamiltonians, $\theta$ will be determined by Eq.\ (\ref{thetai}),
even when $|m_\ell|$ is not small.

It is easy to see that Eq.\ (\ref{thetai}) gives the anticipated
result for our model. Consider first the situation when $\epsilon>6t>0$
which according to Table I is a trivial insulator. In this case all 8
masses $m_\ell$ are positive. Since four of the products
$v_x^\ell v_y^\ell v_z^\ell$ are positive and four are negative Eq.\
(\ref{thetai}) gives $\theta=0$, as expected. Now suppose
we tune $\epsilon$ so that its value drops below $6t$. This reverses
the sign of a single Dirac mass at $\ell={(0,0,0)}$. The corresponding
$\theta^\ell$ also reverses sign and we obtain $\theta=\pi$, as
expected for a STI. For all other cases the values of $\theta$ are listed
in Table I. These results confirm the one-to-one correspondence
between the $Z_2$ invariant $\nu_0$ and the axion parameter $\theta$
expressed by $\theta=\pi\nu_0$, as expected on very general
grounds.\cite{zhang33}

%%%%%%%%%%%%%%%%%%%%%%%%%%%%%%%%%%%%%%%%%%%%%%%
\section{Charge bound to a monopole}

We now consider a magnetic monopole in the interior of a STI. We model
this situation by the Hamiltonian $H$ defined by Eqs.\ (\ref{hso}) and
(\ref{hcd}) with a monopole positioned at the center of a cubic unit
cell. The magnetic field of the monopole couples to both the
electron charge and the electron spin through the orbital and 
Zeeman couplings, respectively. The form of the orbital coupling is
dictated by gauge invariance and is thus universal; in our lattice
model it is implemented by the Peierls substitution, which attaches
factors $e^{i\vartheta_{ij}}$ to all hopping terms connecting sites
$i$ and $j$. Here $\vartheta_{ij}=(2\pi/\Phi_0)\int_i^j\bA\cdot d{\bf l}$ and $\bA$ is
the magnetic vector potential. The Zeeman coupling is of the form
$-g\mu_B \bB\cdot{\bf S}/\hbar$ where $\mu_B=e\hbar/2m_ec$ is the Bohr
magneton and ${\bf S}$ denotes the electron spin. For free electrons
$g$ is close to 2 but in solids the effective $g$ can be substantially
larger. The Zeeman coupling thus leads to an additional term in the
Hamiltonian,
\begin{equation}\label{hz} H_Z=-g\mu_B{1\over 2}\sum_j \bB_j\cdot
\left(\Psi_{j}^\dagger{\bm\sigma}\Psi_{j} \right),
\end{equation}
where $\bB_j$ is the magnetic field at site $j$ of the lattice. This
term is non-universal and its importance will depend on the ratio of
$g\mu_B|\bB|$ to the other relevant energy scales in the Hamiltonian
set by $\lambda$, $\epsilon$ and $t$.

We solve the Hamiltonian $H=H_{\rm SO}+H_{cd}+H_Z$ in
a cube containing $L^3$ sites by exact numerical diagonalization. 
The monopole is placed inside the central unit
cell (at the origin), so that the  magnetic field of the
monopole is ${\bB=(\Phi_{0}/4\pi r^{2}) \hat{r}}$. We choose a gauge in 
which the system retains the four-fold rotational
symmetry around the $z$ axis,\cite{shnir1}  
${\bA = -\Phi_{0}(1+\cos \theta) \nabla \varphi}$, with
$(\theta,\varphi)$ the spherical angles.  
Exploiting this symmetry we are able to simulate system sizes up to $L=20$,
which requires diagonalizing a complex valued Hermitian matrix of size
${1\over 4}(4\times20^3) = 8,000$. In order to calculate the charge
density at half filling we require knowledge of all the occupied
eigenstates.

\begin{figure*}[t]
\includegraphics[width= 15cm]{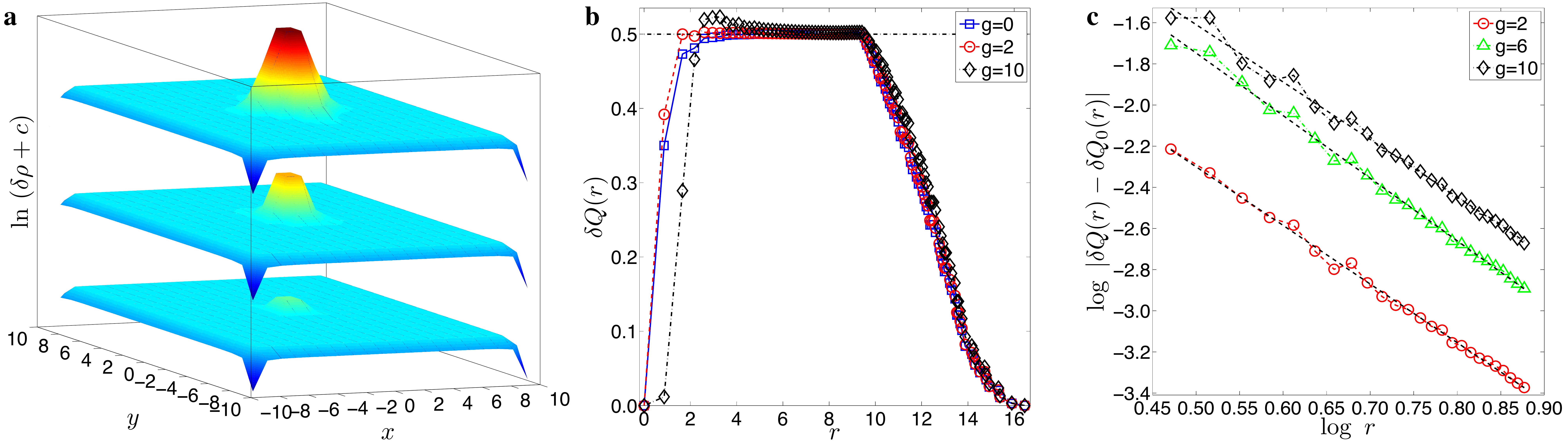}
\caption{(Color online) Charge density in our model TI on the
cube-shaped lattice with $20^{3}$ sites with a unit monopole at its
center, with parameters $t=\lambda$, $\epsilon=4t$, leading to a bulk gap $\Delta=4t$. (a) Charge density
$\delta\rho$ of the three closest layers below the monopole, for
$g=0$. (b) The excess charge $\delta Q(r)$ (in units of $e$) for different
 Zeeman coupling $g$. The knee feature seen at $r=10$ corresponds to the
 radius at which the sphere used to calculate $\delta Q(r)$ first touches the system boundary. (c) Log-log plot of $\delta Q_{g}-\delta Q_{0}$
showing the power-law approach $\sim r^{-\alpha}$ of the accumulated
charge to its assymptotic value of $1/2$. The least-square fit yields
exponents $\alpha=2.85,3.04,2.79$ for $g=2,6,10$, respectively. We attribute the deviations of the numerically 
determined exponent $\alpha$ from the expected value of 3 to the finite size effect.}
\label{fig_monopole}
\end{figure*}

We diagonalize the Hamiltonian, once with the magnetic monopole and
once without, obtaining charge densities $\rho_{1}$ and $\rho_{0}$,
respectively. The monopole-induced charge density $\delta\rho =
\rho_{0}-\rho_{1}$ is plotted in Fig.~\ref{fig_monopole}a. To determine the
total  charge bound to the monopole we calculate the excess accumulated charge in a sphere of radius $r$ centered on the monopole, $\delta
Q(r)=\sum_{|\br_i|<r} \delta\rho(\br_i)$. We find  (Fig.~\ref{fig_monopole}b)
that it saturates at $-e/2$ to within 4 significant digits, comparable to the accuracy of our numerics.
For $g=0$ we find two localized zero modes, one at the monopole and
one on the surface. Fractional charge bound to the monopole can be
understood in this case by appealing to the standard arguments\cite{jackiw1,goldstone1}
developed originally to describe charge fractionalization in
polyacetylene.\cite{su1}  Briefly, when a topological defect (such as a domain wall in polyacetylene) produces a localized zero mode inside 
the gap in a particle-hole symmetric system, one can show that the spectral weight of the state contains equal contributions from the valence and the conduction bands. Thus, the valence band shows a net deficit of half a state in the vicinity of the defect. This translates into the defect carrying fractional charge $\pm e/2$, the sign depending on whether the zero mode is empty or occupied.

Like in polyacetylene we find the saturation of charge to be exponential
$\sim\exp{(-r / \xi)}$, where $\xi \sim 1/\Delta$ and $\Delta$ is the
bulk gap. 

When $g>0$, the Zeeman coupling causes changes in the charge distribution
near the monopole but the total accumulated charge remains quantized
at $-e/2$. In this case there are no exact zero modes in the spectrum
and $\delta Q(r)$ approaches $e/2$ as a power law with the exponent close
to $-3$ (Fig.~\ref{fig_monopole}c). 

The power-law dependence can be understood as follows. The Zeeman term
acts as an additional time-reversal breaking field which modifies the
value of axion $\theta$ away from $\pi$ close to the monopole. This
causes non-vanishing $\nabla\theta$ and thus, according to Eq.\
(\ref{GaussLaw}), additional contribution to the effective charge
density. The simplest assumption, $\delta\theta\sim\bB^2$, gives
$\delta\rho \sim \nabla \theta \cdot \bB \sim r^{-7}$ and $\delta Q
\sim r^{-4}$, a power law but with the exponent not quite in agreement
with our numerical simulation. On further reflection one realizes that
in our model $\delta\theta$ cannot be proportional to $\bB^2$ but
rather must be proportional to its gradients. This is because in the
presence of a uniform Zeeman term the system remains
inversion-symmetric.  Inversion symmetry dictates quantized value of
$\theta=0,\pi$ even when $\cal T$ is explicitly
broken.\cite{qi1,essin1} Thus, non-vanishing $\delta\theta$ requires
spatially varying Zeeman field. The simplest assumption satisfying
these requirements is $\nabla\theta\sim \nabla^2\bB$. In the
vicinity of the monopole one finds $\delta\rho \sim \nabla \theta
\cdot \bB \sim r^{-6}$ and $\delta Q \sim r^{-3}$ in agreement with
our numerical results. 

We note that the above considerations are based on the effective axion
action (\ref{ax}) and apply on lengthscales large compared to
$\xi$. The power law tail in the fractional charge distribution for $g>0$
appears on top of a short-lengthscale structure with a roughly exponential profile that is controlled by the 
properties of the microscopic Hamiltonian and is thus non-universal. At
the intermediate lengthscales the interplay of the two contributions can 
give rise to interesting structures such as the peak in $\delta Q(r)$ at $r\simeq 2.5$ seen in Fig. 1b for $g=10$.  

By the same method described above we have investigated spin density
induced by the monopole. We find that there is no net spin
$\langle{\bf S}\rangle$ attached to the monopole. Thus, in addition to
charge fractionalization, a magnetic monopole inserted in a STI
constitutes an example of spin-charge separation in three spatial
dimensions. This is perhaps not surprising in view of the fact that
spin-orbit coupling present in the Hamiltonian (\ref{hso}) breaks the
SU(2) spin symmetry and, as a result, electron spin is not a good
quantum number in the model describing our system.

%%%%%%%%%%%%%%%%%%%%%%%%%%%%%%%%%%%%%%%%%%%%%%%%%%%%%
\section{Proposal for experimental realization}

Although there is no known theoretical principle that prohibits the
existence of fundamental magnetic monopoles in nature,\cite{shnir1}
none have been observed to date despite extensive
searches.\cite{milton1} This null observation has led to a consensus
that fundamental monopoles either do not exist for some heretofore
unknown reason or they are very rare in our part of the universe. In
either case the observed absence of fundamental monopoles poses a
challenge to the idea of experimental verification of the Witten
effect using a STI. At best, one could conceive of a new scheme for
possible {\em detection} of magnetic monopoles using a STI in the role
of a sensor if a convenient way to detect the fractional charge could
be found.

A much more promising avenue for the verification of the Witten
effect is suggested by exploiting {\em emergent} instead of
fundamental monopoles. A classic example of such an emergent behavior
in a crystalline solid is the 2007 theoretical
prediction\cite{sondhi1} and the subsequent experimental
observation\cite{fennell1,morris1,bramwell1} of monopoles in
frustrated magnetic systems called `spin ice', realized in certain
magnetic pyrochlore compounds such as Dy$_2$Ti$_2$O$_7$ or
Ho$_2$Ti$_2$O$_7$. Magnetic monopoles in these systems arise as
elementary excitations above the collective ground state of spins and
the monopole-like magnetic field configuration originates from the
magnetic moments of the constituent spins. In principle, the emergent
monopoles in the spin ice could be used to test the Witten effect if a
compound that is simultaneously a STI {\em and} a spin ice could be
identified. Unfortunately no such material is known at present
although we note that STI behavior has been theoretically predicted to
occur in crystals with the same pyrochlore structure\cite{guo1,pesin1} that
underlies the spin ice behavior. It is thus possible that a suitable
material will be discovered in the future.

Here we focus on a different type of emergent magnetic monopole that
can arise in a thin film STI placed in a uniform external electric
field.  The basic idea and the feasibility of its experimental
realization have been discussed in Ref.\
\onlinecite{seradjeh1}. Following that work we envision the simplest STI with
just one gapless Dirac state per surface and the chemical potential
initially tuned to the neutral point. When a strong enough electric
field is applied perpendicular to the plane of the film the chemical
potential undergoes a shift that is opposite in the two surfaces. This
creates a small electron Fermi surface in one surface and a small hole
Fermi surface in the other. The essence of the
proposal\cite{seradjeh1} lies in the observation that an arbitrarily
weak Coulomb interaction between the surface states
produces an exciton condensate, which may be viewed as a coherent
fluid of electron-hole pairs drawn from the opposite surfaces. Such an
exciton condensate is characterized by a complex scalar order
parameter $\Phi$, which can fluctuate in space and time. In particular
$\Phi=\Phi_0e^{i\chi}$ can contain {\em vortices} -- point-like
topological defects with the phase $\chi$ winding by $\pm 2\pi$
around a vortex. It has been pointed out in Ref.\
\onlinecite{seradjeh1} that to electrons in a STI such a vortex is
indistinguishable from a `planar monopole', i.e.\ a monopole with
magnetic field radiating in the plane of the surface.
\begin{figure*}[t]
\includegraphics[width= 15cm]{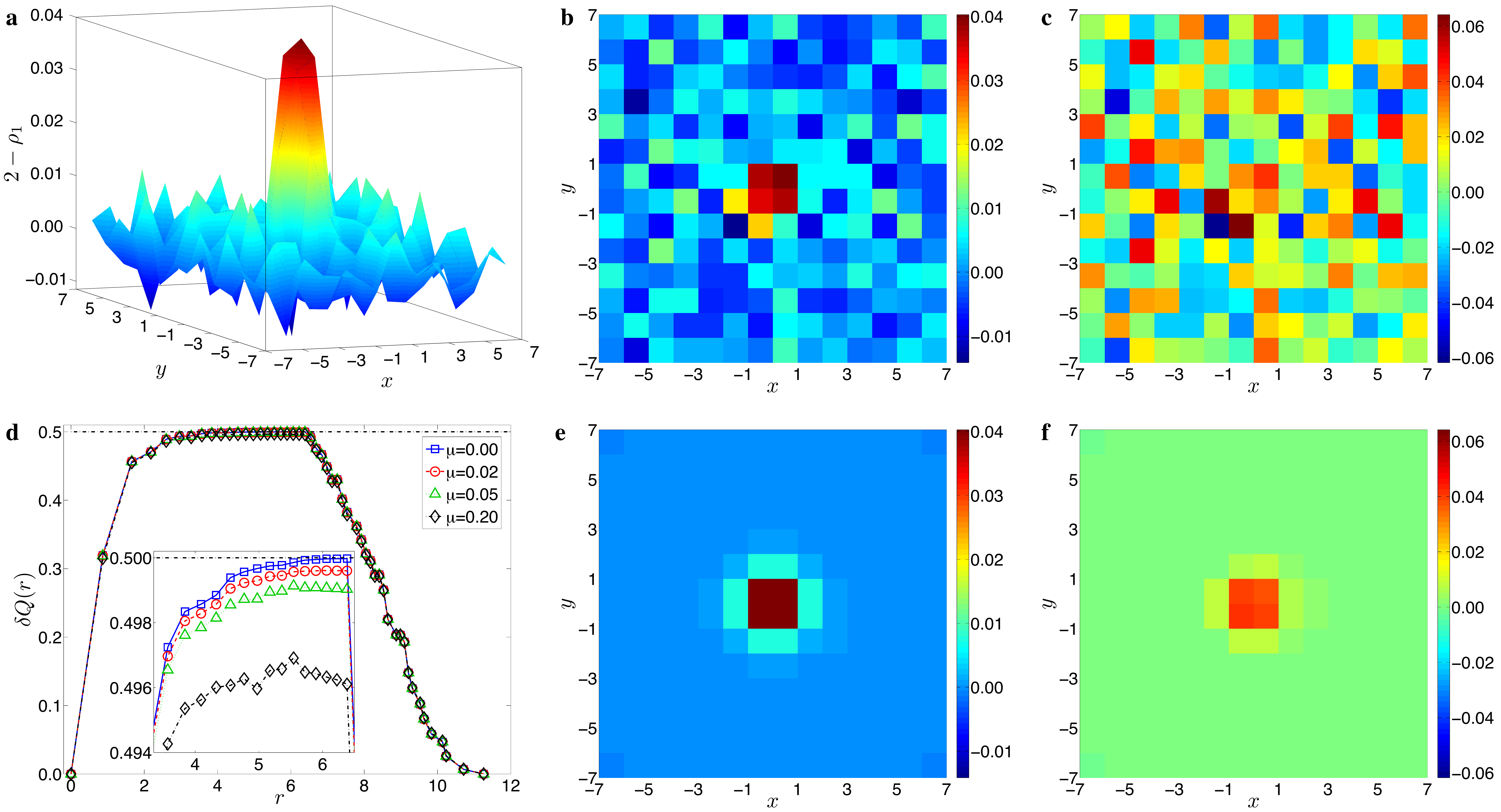}
\caption{(Color online)
A cubic sample of a TI including disorder with a planar unit monopole at its center, 
size $L=14$ and parameters as in Fig.\ ~\ref{fig_monopole}. In
all cases shown we use the same disorder realization but vary its overall
strength parametrized by $\mu$. Panels (a)
and (b) show charge density $2-\rho_{1}$ for the layer just below
the planar monopole for weak disorder $\mu=0.05\Delta$, and (c) larger disorder $\mu=0.20\Delta$. Panels (e) and (f) show the
difference in charge density $\delta\rho=\rho_{0}-\rho_{1}$ for
$\mu/\Delta=0.05,0.20$, respectively, for the same layer. (d) The
excess charge $\delta Q(r)$ (in units of
$e$) for different values of disorder strength $\mu$. The inset shows a close up of the saturation. At this scale a small deviation from the expected asymptotic value $1/2$ that increases with the disorder strength becomes visible. We attribute this deviation to the finite-size effect in our numerical calculation. This identification is supported by the fact that the deviations grow more pronounced for smaller system sizes and close to the surface. Also, it is consistent with the notion that the bound charge is localized on the length scale $\xi\sim 1/\Delta$ which increases as the disorder reduces the spectral gap. 
}
\label{fig_plmonopole}
\end{figure*}

A planar monopole can be viewed as an adiabatic deformation of an
ordinary monopole achieved by flattening the field lines in a
cylindrically symmetric fashion. One expects that the total charge
bound to the monopole via the Witten effect should be insensitive to
such an adiabatic deformation and therefore a vortex in the exciton
condensate should bind fractional charge $\pm e/2$. This indeed has
been argued to happen in Ref.\ \onlinecite{seradjeh1} based on the 
Dirac equation describing the low-energy physics
of the surface states in the presence of the exciton condensate. Here,
taking a more general point of view, we establish the existence of the
fractional charge in such a condensate by studying a planar monopole
embedded inside a STI. Our calculation below does not rely on the
low-energy approximation for the surface states and is insensitive to
the detailed microscopic structure of the condensate. Rather, it
exploits only the most fundamental property of the STI given by its
nontrivial axion response.

In general, the fractional charge is expected to be robust against weak
disorder that does not break TRI. Such a disorder will be present in a real sample and we model it here by adding a term
\begin{equation}
\label{hd} 
H_D = \sum_j \mu_j^+ \Psi_{j}^\dagger\Psi_{j} + \sum_j \mu_j^- \Psi_{j}^\dagger \tau_z \Psi_{j},
\end{equation}
to the Hamiltonian. The first term represents a parity-preserving on-site disorder (independent of the orbital), while the second term is a parity
breaking disorder.

As before we solve the Hamiltonian ${H=H_{\rm SO}+H_{cd}+H_D}$, in a cube
containing $L^{3}$ sites,  by exact numerical diagonalization. The
planar monopole projects an effective magnetic field
$\bB_{\rm eff}=(\Phi_{0}/2\pi r)\delta(z) \hat{r}$ (in cylindrical
coordinates) and the vector potential can be chosen as ${\bA = (\Phi_{0}/2\pi)\varphi\delta(z) \hat{z}}$. The effective field does not couple to electron spin,\cite{seradjeh1} so there is no Zeeman term in this case.  The disorder coefficients $\mu_j^\pm$ are chosen from a Gaussian
distribution with standard deviation $\mu$. Note that the disorder breaks the
four-fold rotational symmetry of the system, so we cannot exploit this symmetry
in this case to efficiently diagonalize the Hamiltonian. Consequently we are
limited to system sizes up to $L=14$. For weak disorder $\mu \ll \Delta$ the charge
bound to the planar monopole remains $-e/2$ (see Fig.~\ref{fig_plmonopole}), and for strong disorder $\mu \gg \Delta$ the charge bound is zero. Remarkably,
even for fairly significant disorder  (such that it generates charge density
fluctuations comparable to the charge density induced by the monopole) 
the difference in charge density $\delta \rho$ shown in Fig.~\ref{fig_plmonopole}f is only weakly affected. 

In the framework of the current proposal the key ingredient required
to produce a monopole-like configuration is the exciton condensate. As
explained in Ref.\ \onlinecite{seradjeh1} it is difficult to reliably
estimate the critical temperature $T_{EC}$ for the formation of the
exciton condensate, but under optimal conditions it should be
higher than it is in bilayer graphene, where the occurrence of this
effect is hotly debated. Once the exciton condensate is formed,
vortices can be nucleated by applying an in-plane magnetic field. Since
the exciton condensate is itself insulating, the main conduction
channel in this situation will be through vortices, each carrying
$-e/2$ charge. Fractional charge of the carriers then can be detected
using established techniques.\cite{goldman1,picciotto1}

%-------------------------------------------------------------------------
\section{Outlook and open questions}
%(written  1-10-2010, Berkeley CA)%

Predicted more than 30 years ago in the context of high-energy physics,
but never before observed in a real or numerical experiment, the Witten
effect is realized in a strong topological insulator. A unit magnetic
monopole inserted in a model STI binds electric charge $-e/2$ in
accordance with the prediction\cite{witten1} and furnishes a rare
example of charge fractionalization and spin-charge separation 
in 3 spatial dimensions. In the
special case when the underlying system possesses particle-hole
symmetry and when the Zeeman term (\ref{hz}) can be neglected, the
appearance of fractional charge follows from the same `zero-mode'
arguments that underlie charge fractionalization in a one-dimensional
system of fermions coupled to a scalar field with a soliton
profile\cite{jackiw1,goldstone1} as realized in dimerized
polyacetylene.\cite{su1} In the more general case when the Zeeman
term or weak disorder are present, there exist no exact zero modes in
the spectrum of electrons yet the fractional charge remains precisely
quantized. This reflects the more subtle topological order that
underlies the axion response of a STI, which is robust against any weak
perturbation that respects time reversal symmetry.\cite{mele1,moore1,qi1,essin1}

An interesting open question is the fate of the Witten effect in the
presence of magnetic disorder. Experimentally, this can be implemented
by adding a small concentration of magnetic ions (such as Fe or Mn)
into a topological insulator. Although at the microscopic level
$\theta$ is no longer quantized in the presence of ${\cal T}$-breaking
perturbations, we expect its effective value, relevant to the physics at
long lengthscales, to remain pinned at $\pi$ as long as the magnetic
moments stay disordered. This is because the net magnetic moment in a
macroscopic region containing many impurities will effectively
vanish. Qualitatively, this suggests that the Witten effect may
survive inclusion of a moderate concentration of magnetic dopants with
randomly oriented moments.  At low temperatures moments may order
ferromagnetically.\cite{yu1} In this case both ${\cal T}$ and the inversion
symmetry are broken (the latter due to the random position of magnetic
dopants) and the effective $\theta$ can acquire an arbitrary value. In
this situation we expect a monopole to still bind fractional charge
according to Eq.\ (\ref{q}) but we leave a detailed study of this case
to future investigation.

Can the Witten effect be observed experimentally in the near
future? We believe that the answer is affirmative. One essential
ingredient, the axion medium, is now widely available in any of
the recently discovered STIs.\cite{cava1,cava2,chen1} If an emergent
monopole can be realized, exploiting the proposed exciton
condensate,\cite{seradjeh1} the spin ice-type
physics,\cite{sondhi1,fennell1,morris1,bramwell1} 
or by some other means, then the
experimental challenge is reduced to designing a suitable method for
the detection of fractional charge bound to the monopole.  The fractional
charge of elementary excitations in fractional quantum Hall fluids has
been previously detected\cite{goldman1,picciotto1} and it should be
possible to adapt these methods to topological insulators. In this
way, studies of crystalline quantum matter with non-trivial
topological properties could help settle one of the enduring
challenges of fundamental physics and provide new insights into
the behavior of electrons placed in unusual situations.

\subsection*{Acknowledgment}
The authors are indebted to I.~Affleck, A.M.~Essin, I.~Garate, H.-M.~Guo, J.E.~Moore, C.~Weeks, X.-L.~Qi, and S.-C.~Zhang for helpful discussions and correspondence. This work was supported by NSERC and CIfAR and was performed, in part, at the Aspen Center for Physics.

%%%%%%%%%%%%%%%%%%%%%%%%%%%%%%%%%%%%%%%%%%%%%%%%%%%%%%%%%%%%%%%%%%%%%%%%%%
\appendix
\section{Evaluation of  $\theta$}

The feature that makes our model Hamiltonian (\ref{hk}) easy to
analyze, is its matrix structure in the combined orbital/spin space,
which consists of four anticommuting $4\times 4$ matrices. Such
matrices are by convention denoted $\gamma_\mu$, $\mu=0,1,2,3$ and
form a representation of the Clifford algebra defined by the
anticommutation relation
$\{\gamma_\mu,\gamma_\nu\}=2\delta_{\mu\nu}$. In terms of these matrices Eq.\
(\ref{hk}) can be written as
\begin{equation}\label{hk2} {\cal H}_\bk=\sum_\mu\gamma_\mu d_\mu(\bk)
\end{equation}
with
$d_\mu(\bk)=(m_\bk,-2\lambda\sin{k_x},-2\lambda\sin{k_y},-2\lambda\sin{k_z})$. The
actual form of our $\gamma$ matrices is apparent by comparing
Eq.(\ref{hk2}) to (\ref{hk}) but our result for $\theta$ is
independent of any particular representation, as long as the matrices
satisfy the requisite commutation relation. In fact in the subsequent
calculations it will be advantageous to use a different representation
of the Clifford algebra, obtained by a uniform rotation by angle
$\pi/2$ around the $\tau_1$ axis: we use $\gamma_0=\tau_1\otimes\id$
and $\gamma_i=\tau_2\otimes\sigma_i$, $i=1,2,3$.

In this representation the two normalized negative-energy eigenstates
of ${\cal H}_\bk$ can be written as
\begin{eqnarray}\label{psi} \psi_1 &=&
(-d_1+id_2,d_3-id_0,0,iE)^T/\sqrt{2}E, \nonumber \\ \psi_2 &=&
(d_3+id_0,d_1+id_2,-iE,0)^T/\sqrt{2}E,
\end{eqnarray}
where $E=(\sum_\mu d_\mu^2)^{1/2}$, and from now on we suppress the
momentum dependence of all quantities. The eigenstates $\psi_1$ and
$\psi_2$ above are degenerate and we are thus free to choose any
(orthogonal) linear combination of these. Such a change of basis
corresponds to a gauge transformation on $\cA$. When evaluating
$\theta$ using Eq.\ (\ref{theta}) one must keep in mind that the
integrand is not gauge invariant while the integral taken over the BZ
is gauge invariant modulo 4$\pi^2$. This property reflects the $Z_2$
nature of the topological invariant $\nu_0$ that underlies the physics
of a STI.

Using the eigenstates given in Eq.\ (\ref{psi}) we find the Berry
connection to be of the form $\cA_i={\bf n}_i\cdot{\bm \sigma}$, where
${\bm \sigma}$ is a vector of Pauli matrices and the components of
vector ${\bf n}_i$ read
\begin{eqnarray}\label{n} n_{1i} &=&
D_0\partial_iD_1-D_1\partial_iD_0+D_3\partial_iD_2-D_2\partial_iD_3,
\nonumber \\ n_{2i} &=&
D_3\partial_iD_1-D_1\partial_iD_3+D_2\partial_iD_0-D_0\partial_iD_2,
\nonumber \\ n_{3i} &=&
D_2\partial_iD_1-D_1\partial_iD_2+D_0\partial_iD_3-D_3\partial_iD_0,
\nonumber
\end{eqnarray}
with $D_\mu=d_\mu/\sqrt{2}E$. After substituting these into Eq.\
(\ref{theta}) a tedious but ultimately straightforward calculation
leads to the expression
\begin{equation}\label{thetad} \theta=-{1\over 2\pi}\int_{\rm
BZ}d^3k\epsilon^{\alpha\beta\mu\nu}{d_\alpha\partial_1
d_\beta\partial_2 d_\mu\partial_3 d_\nu \over E^4}.
\end{equation}
Using the values of $d_\mu$ for our model given below Eq.\ (\ref{hk2}),
one obtains a complicated integrand in terms of trigonometric
functions. Although the value of the integral is guaranteed to be
either 0 or $\pi$ in practice it is not obvious how to perform the
required three-dimensional integration. However, it is clear from the
structure of the integrand in (\ref{thetad}) that in the limit
$|m_\bk|\ll\lambda$ contributions to the integral come only from the
vicinity of the 8 Dirac points. We evaluate these 8 contributions
separately by linearizing $d$'s as $d_\mu^\ell(\bk)=(m_\ell,v_x^\ell
k_x, v_y^\ell k_y, v_z^\ell k_z)$ and obtain
\begin{equation}\label{thetadlin} \theta^\ell=-{1\over 2\pi}\int
d^3k{m_\ell v_x^\ell v_y^\ell v_z^\ell \over
(4\lambda^2k^2+m_\ell^2)^2}.
\end{equation}
An elementary evaluation then yields the result quoted in Eq.\
(\ref{thetai}).

\end{document}